\documentclass[12pt]{article}

\usepackage{amsmath}
\usepackage{amssymb}
\usepackage{verbatim}

\makeatletter

   \@addtoreset{equation}{section}
\makeatother

\newcommand{\be}{\begin{equation}}
\newcommand{\ee}{\end{equation}}
\newcommand{\bea}{\begin{eqnarray}}
\newcommand{\eea}{\end{eqnarray}}

\begin{document}

\vskip 12mm

\begin{center}

{\Large \bf Entanglement Entropy from String Field Theory (and  a Higher-Spin Example)}
\vskip 10mm
{ \large  Dimitri Polyakov$^{a,b,}$\footnote{email:polyakov@scu.edu.cn;polyakov@sogang.ac.kr}
}

\vskip 8mm
$^{a}$ {\it  Center for Theoretical Physics, College of Physical Science and Technology}\\
{\it  Sichuan University, Chengdu 6100064, China}\\
\vskip 2mm

$^{b}$ {\it Institute of Information Transmission Problems (IITP)}\\
{\it  Bolshoi Karetny per. 19/1, Moscow 127994, Russia}

\end{center}

\vskip 15mm

\begin{abstract}

We study the new class of solutions in linearized open string field theory
(OSFT)
involving higher-spin modes. Unlike the elementary OSFT solutions 
(on-shell vertex operators) that, acting on a vacuum, define wavefunctions
of pure states (e.g. a tachyon), the solutions that we describe
correspond to the reduced density matrices which eigenvalues describe the entanglement
between higher-spin modes with different spin values.
We compute the entanglement entropy on these OSFT
solutions, and the answer is expressed in terms of converging
series in inverse weighted partition  numbers.
In the  case of $D$-dimensional bosonic  string theory, the entanglement entropy
of spin $1$ subsystem and the system of all the spin values is given by
$D{\log{\lambda_0}}+{D\over{\lambda_0}}\sum_{N=3}^\infty{{|\beta(N)|}\over{\lambda(N)}}
{\log{({{\lambda(N)}\over{|\beta(N)|}})}}$,
where $\lambda(N)$ is the weighted number of partitions of $N$,
$\beta(N)={{(N-1)\zeta(3)-\zeta(2)}\over{(N-1)^4}}$
and $\lambda_0=\sum_{N=1}^{\infty}{{\beta(N)}\over{\lambda(N)}}$.
The  first term, $D{\log{\lambda_0}}$,
represents the entanglement swapping between string vacuum and string excitations.
 We generalize this result 
to obtain the entanglement for a subsystem of a given spin $s$ in 
a given space-time dimension. We also discuss how 
open string field theory may be used to study the entanglement of systems
 other than higher spin excitations in string theory.

\end{abstract}

\vskip 12mm

\setcounter{footnote}{0}

\section{\bf Introduction}

The concept of entanglement  has recently attracted a lot of interest due to its relevance
 to  building an interface between information theory, quantum gravity
and strongly coupled field theories, including some condensed matter systems
(a very incomplete and subjective list of conceptual works on the subject, both classical  and recent, includes,
but is not limited to
 \cite{ein, cardy, malda, myers, raam, raams, ryu, hart, faulkner, bal, bin, dong, bino}, as well as
many other remarkable
works on the subject)
The entanglement also appears to be a crucial ingredient in our attempts to understand the microscopic structure of
space and time, the emergence  of space-time geometry and to understand gravity in the context of quantum mechanics 
(e.g. see \cite{raam, raams}).
Entanglement entropy  is a particularly important quantity characterizing quantum-mechanical or field-theoretic
subsystems in mixed states that interact with some other systems and are described by reduced 
density matrices (rather than wavefunctions). In some cases, it can be measured experimentally for  certain
 systems,  such as ultra-cold atoms or entangled photons, leading to some fascinating observations,
such as quantum non-locality in space and time \cite{xiao, meg}.
One particularly interesting example of a system where the entanglement occurs naturally 
is the one of higher-spin fields,
which by themselves constitute an important ingredient of gauge-to gravity correspondence 
and have been an subject of
 a deep interest and investigation over recent years 
(some conceptual works on the subject include, but are not limited to \cite{fronsdal, fvf, fvs, vmas, bek, kleb}).
 It is well-known, from the structure of the higher-spin symmetries,
that it is impossible to consistently truncate these theories at spin values greater than $2$: for example, a
commutator of two spin $3$ currents would inevitably contain a contribution with spin $4$, and so on.

From the point of view of quantum mechanics this means that a system of particles with a given spin $s\geq{3}$ 
cannot be described by a wavefunction, but has to be a part of a density matrix which structure
reflects the entanglement  of this system with other higher-spin fields.
In particular, this raises natural questions about higher-spin modes appearing in string theory:
constructing on-shell vertex operators for massive higher-spin states is straightforward in string-theoretic
formalism, but each of these on-shell operators acting on the vacuum defines certain pure quantum-mechanical state,
so neither non-locality of interactions, nor entanglement are obvious in the on-shell approach.
In our work, we particularly address this question by describing the off-shell solution in open string field theory,
which particularly reflects the entanglement between the higher-spiun modes.
In fact, it turns out that even lower-spin system with $s<3$ is entangled with the higher spins, although
classically the higher-spin algebra can be truncated to the lower-spin currents, generating the 
underlying space-time isometries (e.g. $AdS$ isometry algebra).
This by itself makes higher-spin systems  an instructive example to study entanglement.

In general, the entanglement entropy is hard to compute in quantum field theory since
the computation involves complicated functional integrals. For example, to calculate the $n$'th Renyi 
entropy in 2d conformal field theory  one has to evaluate the partition function on $n$ glued copies of a Riemann
surface which in general is highly non-trivial.

String field theory, on the other hand,  can be regarded a natural framework to  explore the entanglement - in  particular, 
in the case of higher-spin modes in string theory.
 First of all, a string field in the second-quantized theory is by definition
an expansion in operators, with the spin being  a natural expansion parameter. 
Also, the underlying equations for higher-spin fields
 have a form similar to Vasiliev's equations \cite{witsft, witsfts, berkf}
making components of string fields the objects reminiscent of the differential forms in the higher-spin equations. 
At the same time, string 
field theory is background independent: shifting a string field by an analytic solution of the
equations of motion leaves the form of the equations invariant, with the new nilpotent BRST charge, which
cohomology defines an on-shell theory in a new background. The space-time geometry is therefore emergent
 in string field theory, making it a natural interface to test the entanglement, in the context
of the recent ideas relating the  space-time origin to quantum information
 \cite{raam, raams}.

Finding analytic solutions to the string field theory equations is generally hard and , despite some progress
over recent years, very limited number of non-trivial solutions is known. There exists, however, a well-known 
class of elementary solutions of these equations ($Q\Psi+\Psi\star\Psi=0$). For example,
any on-shell vertex operator in string theory having the form $V\sim{c}P(\partial{X},\partial^2X,...)e^{ikX}\varphi(k)$
 solves the linearized equation $Q\Psi=0$ if V is a dimension 0 primary field - both the 
linearized 
(here $P$ is polynomial in derivatives of the target space field $X_m(z)$ in bosonic string theory).
Each of these solutions, acting on a vacuum, defines a physical state in open string theory.
From the quantum-mechanical point of view these are the pure states, with $V$ defining the wavefunction.
From the space-time point of view, each of these states belongs to some irreducible representation
of the the Lorentz group and is labelled by eigenvalues of Casimir operators of the Poincare algebra in space-time.
As we show in this work, apart from these elementary solutions there exists a class of SFT solutions
defining mixed states on ensemble of wavefunctions;  in fact, these solutions already appear
at the linearized level which we discuss in our work.
This new class of solutions describes the states that are not Casimir eigenvectors and have no definite mass or spin;
 In general, these solutions
have the form of infinite formal series in higher-spin operators with different spins and masses, and with the expansion coefficients
describing the entanglement between the sectors with different spins. 
So if we understand the string states with definite spins and masses as pure states, the 
BRST cohomology solutions that we present
in our work  describe the nontrivial {\it ensembles} consisting of the above atates, i.e. the mixed states from the quantum-mechanical 
point of view. 

The  entanglement entropy can therefore be defined and 
computed on these solutions and, despite the complexity of the correlation functions involved, the
final answer turns out to be finite and surprisingly simple: the entropies are expressed in terms of converging
series involving the partition numbers for the restricted partitions, with the restriction details
depending on values of the entangled spins. This entropy is ``classical'' in a sense
that it is defined on the solution of the SFT equation of motion, which is classical from the second-quantized point of view.
 In case of 
the entanglement of spin one subsystem with the higher-spin system  the answer
is particularly simple and instructive; thus for large $N$ the contribution of a spin $N$ subsystem to the entanglement
entropy of the spin 1 subsystem is, in the leading order,
given by

\begin{equation}
S_{1-N}\sim{{e^{-\alpha{\sqrt{N}}}}}
\end{equation}

where $\alpha$ is constant which can be evaluated  asymptotically.

In case of the entanglement of a spin $s$ subsystem with the rest-of-the-spins system
the structure of partitions involved becomes more complicated but still can be computed explicitly;
in particular we expect that consistency conditions
for the partial entanglements may lead to new non-trivial identities in number theory.
In this paper, we limit this question to the discussion section, 
leaving the  details for the future work.
The rest of the paper is organized as follows.

In the Section 2 we  discuss the simplest example of the 
mixed state type solution appearing
in the linearized bosonic open string field theory, describing the
entanglement of the lowest spin 1 subsystem to
the system including all the tower of the higher spins. The density matrix,
as well as the entanglement entropy are expressed in terms of convergent series
in the inverse weighted partition numbers $\lambda^{-1}(N)$ of integers $N>0$.
To compute the OSFT correlators, relevant to the  solution, we use the
singularization transformation (described in the paper), which is the conformal transformation  
making it possible to express the reduced density matrix and the entropy in terms 
of generalized Schwarzians and 
ordered Bell numbers that in turn can be simplified and expressed in terms
of simple combinations of the partitions.

In Section 3 we generalize the calculation  to obtain solutions describing the entanglement
of a given spin $s$   subsystem with the ensemble containing other higher-spin modes. 
It turns out that, for $s>1$ it is more convenient to use the framework of
 RNS superstring theory \cite{berkf} rather than that bosonic theory. Namely, we identify
 the analytic solutions at superconformal ghost number $s\geq{-3}$
in the cohomological gauge (used instead of the standard gauge ${\eta_0}\Psi=0$)
with mixed states with the reduced  density matrix describing
 the entanglement of the spin $s$ field with the ensemble.
The result is again expressed in terms of relatively simple
convergent series involving restricted partition numbers, with the character of
the restrictions  depending on $s$.

In the concluding section, we discuss physical implications of our results
for the interplays between string dynamics and
entanglement, as well as their generalizations for systems beyond higher spins.

\section{\bf Bosonic SFT and lower-higher spin entanglement}

Consider open  bosonic string field theory equation of motion: 
\begin{equation}
Q\Psi+\Psi\star\Psi=0
\end{equation}
and its linearization 
\begin{equation}
Q\Psi=0 
\end{equation}
where $\star$ is conformal transformation
putting worldsheets of interacting strings on wedges of a disc,
\begin{equation}
Q=\oint{{dz}\over{2i\pi}}{\lbrack}-{1\over2}c\partial{X_m}\partial{X^m}(z)+bc\partial{c}(z)\rbrack
\end{equation}
is the BRST charge in bosonic string theory, 
skipping the Liouville terms (that ensure the overall nilpotence of $Q$
in non-critical space-time dimensions,
but that shall play no role in our calculations). 
$X^m;m=1,...D$ are the target space coordinates,
$b,c$ are fermionic reparametrization ghosts.
The equation (2.2) has a class of elementary solutions having the form
\begin{equation}
\Psi_0=\sum_{i}cV_i(X(z))\varphi_i(p)
\end{equation}
where $V_i$ are  primary fields of dimension 1 and ghost number 0
(so that $\Psi_0$ is a primary of ghost number 1 and conformal dimension 0).
In this case, $\Psi_0$ simply defines the open string spectrum in the unperturbed theory, modulo gauge transformations.
Each term in the sum (2.4) then defines a physical operator in open string theory which, acting
on a vacuum, defines a wavefunction of such a string mode in space-time. From the
quantum-mechanical point of view all such states are the pure states, with their wavefunctions satisfying 
the low-energy effective action's equations of motion (e.g. a Klein-Gordon equation for a tachyon, solving the
linearized equation (2.2)).
Apart from this class of elementary solutions, the known examples of nontrivial solutions to the full cubic OSFT
equation (2.1) are few since in general the conformal transformations induced
 by the star product
act on $\Psi$ in a highly nontrivial way. One remarkable example of such a solution is the one found by Schnabl \cite{schnablf},
describing the background with nonperturbative configuration of a tachyon potential
(in some sense, the Schnabl's solution can be thought of as a ``nonperturbative tachyon vertex operator at zero momentum'').
One may wonder, however, if the linearized equation (2.2) admits any nontrivial solutions too, other than 
the elementary class (2.4). It turns out that the nontrivial solutions do exist at the linearized level, and they
describe the mixed states related to the entanglement of different spin modes in
open string theory.
Below we shall describe these solutions and compute the higher-spin entanglement entropy for such solutions.
The answer for the density matrix and for the entropy turns out to be remarkably simple, despite the seeming complexity of the correlators
involved. For simplicity, let us start from the $D=1$ case, 
which will be straightforward to generalize to higher space-time dimensions.
Consider a general ghost number one string field, with the Siegel gauge 
constraint 
\begin{equation}
b_0\Psi=0
\end{equation}
More specifically, consider the string field in the Siegel gauge with gost number 1 and with the following expansion in infinite formal series
in derivatives of $X$:
\begin{eqnarray}
\Psi_0=c\sum_{N=1}^\infty\sum_{p=1}^N\sum_{N|n_1...n_p}
\alpha_{n_1...n_p}{{\partial^{n_1}X}\over{n_1!}}...{{\partial^{n_p}X}\over{n_p!}}
\end{eqnarray}
where $\sum_{N|n_1...n_p}$ stands for the summation over ordered length $p$ partitions of N:
\begin{eqnarray}
N=n_1+...+n_p
\nonumber \\
n_1\geq{n_2}...\geq{n_p}>0
\end{eqnarray}
and $\alpha_{n_1...m_p}$ are some coefficients.
The numbers $N$ and $p$ are thus useful parameters of such an expansion;
although not directly related to higher-spin currents in space-time in $D=1$, in higher space-time
dimensions
 conformal dimension (worldsheet spin) $N$ of a string field component 
actually can be related 
to the space-time spin $N$ of the component,
with the contributions from different $p$ looking like ``Stueckelberg-like'' terms.
It is therefore convenient to cast $\Psi_0$ as
\begin{eqnarray}
\Psi_0\equiv{c}\sum_{N=1}^\infty\sum_{p=1}^{N}\Psi_0^{(N;p)}
\end{eqnarray}

Our initial goal will be to find the choice of the coefficients $\alpha$ for which
$\Psi_0$ is the analytic solution of the linearized equation (2.2). In practice, it is convenient
to use the following definition: we shall define $\Psi_0$ as the solution of the equation
$Q\Psi_0=0$ if
\begin{equation}
<<Q\Psi_0,\Psi>>\equiv<Q\Psi_0(0)I\circ\Psi(0)>=0
\end{equation}
for {any} string field $\Psi$ ( {\bf  not} necessarily in the Siegel's gauge) 
Since $\Psi$ is arbitrary, this identity, once true for any two-point correlator,
will also be true for the insertion of $Q\Psi_0$ into any other SFT correlator,
due to the closedness of the full operator algebra in CFT, which is equivalent
to the statement that $Q\Psi_0$ vanishes identically.
Here the double brackets stand for the standard OSFT correlator and
the conformal transformation $I(z)=-{1\over{z}}$
maps $\Psi$ to infinity.

Let us start with evaluating $Q\Psi_0$.
Simple calculation gives:
\begin{eqnarray}
Q(c\sum_{N,p}\Psi_0^{N;p})=\sum_{N,p}(N-1)\partial{c}c
\Psi_0^{N;p}
\nonumber \\
+\sum_{N,p}\sum_{N|n_1...n_p}\sum_{j=1}^p\sum_{k=2}^{n_j}{{\partial^{k}cc}\over{k!}}{{\alpha_{n_1...n_p}\partial^{n_1}X
...\partial^{n_{j-1}}X\partial^{n_j-k}X\partial^{n_{j+1}}X...{\partial{n_p}}X}\over{n_1!...n_{j-1}!(n_j-k)!n_{j+1}!...n_p!}}
\nonumber \\
+\sum_{N,p}
\sum_{N|n_1...n_p}
\sum_{1\leq{i}<j\leq{p}}{{\partial^{n_i+n_j+1}cc}\over{(n_i+n_j+1)!}}
\nonumber \\
\times
{{\alpha_{n_1...n_p}\partial^{n_1}X...
\partial^{n_{i-1}}X\partial^{n_{i+1}}X...\partial^{n_{j-1}}X\partial^{n_{j+1}}X...\partial^{n_p}X}\over
{n_1!...n_{j-1}!n_{i+1}!....n_{j-1}!n_{j+1}!...n_p!}}
\end{eqnarray}
It can be shown, however that,  with $\Psi_0$ having the ghost structure (2.6), (2.8)  only the first term
in $Q\Psi_0$ contributes
to the correlator $<<Q\Psi_0\Psi>>$.
To prove this, note that  $Q\Psi_0$ has ghost number 2, so  the only $\Psi$  components
contributing to the correlator are those having ghost number 1.
The operators having ghost number 1 in general have the ghost part proportional to 
$$:\sim\partial^{m_1}b...\partial^{m_r}b\partial^{n_1}c...\partial^{n_{r+1}}c:\sim{G(\partial\sigma,\partial^2\sigma,...)}e^\sigma$$
where $m_j,n_j$ are non-negative integers and $G$ is some polynomial in derivatives of $\sigma$, with the standard
bosonization relations: $b=e^{-\sigma};c=e^\sigma$. First of all, it is clear that only the terms with $r=0$ or $1$ can contribute
(otherwise there would be $b$-fields left with no contractions). Let us first check our claim for $r=0,n_1=0$ and then generalize it to the arbitrary case. 
 In the case of $r=0,n_1=0$ ($\Psi$-field prioportional to the $c$-ghost)
(ghost number zero or contain powers of the $b$-ghost) the ghost part of the correlator
 $<<Q\Psi_0\Psi>>$ has the form $<\partial^k{c}c(0)I{\circ}c(0)>$.
It is then easy to check that the only nonzero correlator is the one 
at $k=1$. Indeed, write 
$I\circ{c}=({{dI}\over{dz}})^{-1}|_{z=0}c(\infty)={\lim_{w\rightarrow{\infty}}}w^{-2}c(w)$. 
Then, at  $k=1$,

\begin{eqnarray}
<\partial{c}c(0)I{\circ}c(0)>={\lim_{w\rightarrow\infty}}w^{-2}<\partial{c}c(0)c(w)>
={\lim_{w\rightarrow{\infty}}}w^{-2}w^2=1
\end{eqnarray}
At $k=2$,
\begin{equation}
<\partial^2{c}c(0)I{\circ}c(0)>={\lim_{w\rightarrow\infty}}w^{-2}<\partial^2{c}c(0)c(w)>
={\lim_{w\rightarrow{\infty}}}w^{-2}(-{2{w}})=0
\end{equation}
For higher $k>2$ the ghost correlators vanish identically; that is, using the bozonized expression
$c=e^\sigma$ we write $\partial^k{c}c=B^{(k)}_\sigma{e^\sigma}$ where 
$B^{(k)}_\sigma$ is the degree $k$ Bell polynomial in derivatives of $\sigma$; its OPE
with $e^\sigma$ has the form: 
\begin{equation}
B^{(k)}_\sigma(z)e^\sigma(w)=(z-w)^{-1}
k:B^{(k-1)}_\sigma(z)e^\sigma:(w)+O(z-w)^0,
\end{equation} 
so 
\begin{equation}
:\partial^kcc:={k}:B^{(k-1)}_\sigma{e^{2\sigma}}:
\end{equation}
As it is clear from the OPE (2.13), for $k>2$ the polynomial 
$:B^{(k-1)}_\sigma:(0)$ cannot fully contract with the $c$-ghost at infinity
and all such correlators vanish identically.
This constitutes the proof that only the terms proportional to
$N\partial{c}c\Psi_0^{(N,p)}$ ($k=1$)  in $Q\Psi_0$ contribute to the correlator
$<<Q\Psi_0\Psi>>$ with the components of $\Psi$ satisfying $r=0,n_1=0$.
Now let us show that, once this is true  for $r=0,n_1=0$, this is also true for
arbitrary components of $\Psi$. For the reasons pointed out above, it is sufficient to show  that this is
 the case for $r=1$,i.e. for the components of $\Psi$ with the ghost structure $\sim:\partial^{m_1}b\partial^{n_1}c\partial^{n_2}c:$. 
First of all, note that, since the correlator $<<\partial^k{c}c(0) I\circ{c}(0)>>=0$ on the half-plane for $k=0$, it also vanishes under
any conformal transformation: $z\rightarrow{f(z)}$ of the half-plane. Now let us consider the half-plane correlator
$<<\partial^{k}c{c}(0)(I\circ({\partial^{m_1}b\partial^{n_1}c\partial^{n_2}c})(w\rightarrow\infty)>>$ 
(for the certainty, on the upper half-plane) and apply the conformal transformation ${z}{\rightarrow}f(z)=e^{iz}$.
This transformation is well-defined everywhere on the upper half-plane (including the real axis) and vanishes exponentially fast at infinity. Under this transformation,
the $:\partial^{m_1}b\partial^{n_1}c\partial^{n_2}c(z):$ operators transform as
\begin{eqnarray}
:\partial^{m_1}b\partial^{n_1}c\partial^{n_2}c:(w\rightarrow\infty)\equiv{:H(\partial\sigma,\partial^2\sigma,...)e^\sigma:}(w\rightarrow\infty)
\nonumber \\
\rightarrow
{\lim_{w\rightarrow\infty}}{\lbrace}S(m_1|n_1,n_2)(e^{iw};w)c(w)+O(e^{iw})\rbrace
\nonumber
\end{eqnarray}
where we skipped the terms of orders of $e^{iw}$ and higher (suppressed exponentially when $w$ is taken to infinity)
and $S(m_1|n_1,n_2)(e^{iw};w)$ is the generalized Schwarzian of the conformal transformation $z\rightarrow{e^{iz}}$
of the upper half-plane, appearing as a result of the regularization of the internal singularities in 
operator products between the derivatives between of the $b$ and $c$-ghosts. For the exponential conformal transformation
of the half-plane $S(m_1|n_1,n_2)(e^{iw};w)$ are constant numbers that do not depend on $w$ (see below for the 
discussion of some essential properties of the generalized Schwarzians).
For this reason, the correlator $\lim_{w\rightarrow\infty}<\partial^k{c}c(0)\partial^{m_1}b\partial^{n_1}c\partial^{n_2}c(w)>$,
computed on the Riemann surface as the  result of the conformal transformation of the upper half-plane,
is proportional to the correlator $\lim_{w\rightarrow\infty}<\partial^k{c}c(0)c(w)>$ on the same Riemann surface
(with the coefficient given by  constant generalized Schwarzian factor) and  therefore vanishes  for $k>1$.
This constitutes the proof that only the terms proportional to
$N\partial{c}c\Psi_0^{(N,p)}$ need to be considered in $Q\Psi_0$.
We are now prepared to analyze the correlator
$<Q\Psi_0(0)I\circ\Psi(0)>$ for 
$\Psi_0$ of the form (2.6) and 
an arbitrary string field $\Psi$. 

The string fields of this correlator are located on the halfplane's boundary;
the crucial next step to compute the correlator is the conformal transformation of the half-plane:
\begin{equation}
z\rightarrow{f(z)}=e^{iz}
\end{equation}
taking the upper half-plane to compact Riemann surface,
with $Q\Psi_0$ taken from zero to 1 and $\Psi$ from infinity to zero.
This conformal transformation (which we will also refer to as the ``singularization transformation''
) maps the  upper half-plane to a compact Riemann surface which we shall call the ``singularoid''.
 
Consider the behavior of the $<<Q\Psi_0,\Psi>>$ correlator under such a conformal map.
For that, one crucial relation that we shall need is
the transformation law of the $:\partial^{n_1}X\partial^{n_2}X:(z)$-operator
under $z\rightarrow{f(z)}$, given by
\begin{eqnarray}
{1\over{n_1!n_2!}}:\partial^{n_1}{X}\partial^{n_2}X:(z)\rightarrow
\nonumber \\
{1\over{n_1!n_2!}}\sum_{k_1=1}^{n_1}\sum_{k_2=1}^{n_2}B_{n_1|k_1}(f(z);z)B_{n_2|k_2}(f(z);z)
:\partial^{k_1}{X}\partial^{k_2}{X}:(f(z))
\nonumber \\
+S_{n_1|n_2}(f(z);z)
\end{eqnarray}
where
$B_{n|k}$ are the incomplete Bell polynomials in the $z$-derivatives of $f$.
The general definition of $B_{n|k}$ is:
\begin{eqnarray}
B_{n|k}(g_1,...g_{n-k+1})=n!\sum{1\over{p_1!...p_{n-k+1}!}}
{({{g_1}\over{1!}})^{p_1}}...
{({{g_{n-k+1}}\over{(n-k+1)!}})^{p_{n-k+1}}}
\end{eqnarray}
with the sum taken over all the non-negative $p_1,...p_{n-k+1}$
satisfying
\begin{eqnarray}
p_1+...+p_{n-k+1}=k
\nonumber \\
p_1+2p_2+...+(n-k+1)p_{n-k+1}=n
\nonumber
\end{eqnarray}
In particular,
the incomplete Bell polynomials $B_{n|k}(f;z)$  
 in  the derivatives 
(or the expansion coefficients) of $f(z)$,
are given by
$g_k=\partial_z^{k}f(z)\equiv{{d^kf}\over{dz^k}}$
(although the  partial derivative sign is not necessary, we keep it to shorten our notations)
or equivalently

\begin{eqnarray}
B_{n|k}(f(z);z)=n!\sum_{n|n_1...n_k}{{\partial^{n_1}f(z)...\partial^{n_k}f(z)}
\over{n_1!...n_k!q(n_1)!...q(n_k)!}}
\nonumber
\end{eqnarray}
with the sum ${n|n_1...n_k}$ taken 
over all ordered $<n_1\geq{n_2}...\geq{n_k}>0$
length $k$ partitions of $n$ and with $q(n_j)$
denoting the multiplicity of  $n_j$ element of the partition
(e.g.  for the partition $7=2+2+3$ we have $q(2)=2,q(3)=1$,
so the appropriate term would read 
$\sim{{\partial^2{f}\partial^2{f}\partial^3{f}}\over{2!2!3!\times{2!1!}}}$.
Then, $S_{n_1|n_2}(f(z);z)$ are the generalized Schwarzians of the conformal transformation,
given by
\begin{eqnarray}
S_{n_1|n_2}(f;z)=
{1\over{n_1!n_2!}}\sum_{k_1=1}^{n_1}\sum_{k_2=1}^{n_2}
\sum_{m_1\geq{0}}\sum_{m_2\geq{0}}\sum_{p\geq{0}}
\sum_{q=1}^p
(-1)^{k_1+m_2+q}2^{-m_1-m_2}(k_1+k_2-1)!
\nonumber \\
\times
{{\partial^{m_1}B_{n_1|k_1}(f(z);z)\partial^{m_2}B_{n_2|k_2}(f(z);z)
B_{p|q}(g_1,...,g_{p-q+1})}
\over{m_1!m_2!p!(f^\prime(z))^{k_1+k_2}}}
\nonumber \\
g_s=2^{-s-1}(1+(-1)^s){{{{d^{s+1}f}\over{dz^{s+1}}}}\over{(s+1)f^\prime(z)}};
s=1,...,p-q+1
\end{eqnarray}
with the sum over the non-negative numbers $m_1,m_2$ and $p$ taken over all the combinations satisfying
$$m_1+m_2+p=k_1+k_2$$
For $n_1=n_2=1$ $S_{1|1}$ becomes the usual Schwarzian derivative (up to the conventional
normalization factor of ${1\over6}$).
Note that the exponential factors proportional to powers of $\sim{e^{iz}}$
cancel out in all the terms of the summation,
so for the conformal transformation that we need, $f(z)=e^{iz}$, the generalized  Schwarzians
$S_{n_1|n_2}$ do not depend on $z$ and are constant.
For the conformal transformation under study, $f(z)=e^{iz}$, the value of the  Bell polynomials $B_{n|p}(f(z);z)$ 
and their derivatives at can be expressed in terms of the Stirling numbers of the second kind $S(n;k)$:
\begin{eqnarray}
B_{n|k}(e^{iz};z)=i^nS(n;k)e^{ikz}
\nonumber \\
\partial^p_z{B_{n|k}}(f(z);z)=i^{n+p}k^pS(n;k)e^{ikz}
\end{eqnarray}
and accordingly, for $f(z)=e^{iz}$ the explicit form of the generalized Schwarzians
can be simplified to give:
\begin{eqnarray}
S_{n_1|n_2}(f;z)=
{1\over{n_1!n_2!}}\sum_{k_1=1}^{n_1}\sum_{k_2=1}^{n_2}
\sum_{m_1\geq{0}}\sum_{m_2\geq{0}}\sum_{p\geq{0}}
\sum_{q=1}^p
(-1)^{k_1+m_2+q}2^{-k_1-k_2}(k_1+k_2-1)!
\nonumber \\
\times
{{i^{-p}S(n_1,k_1)S(n_2,k_2)k_1^{m_1}k_2^{m_2}
B_{p|q}(g_1,...,g_{p-q+1})}
\over{m_1!m_2!p!}}
\nonumber \\
g_s={{2^{-s}cos({{\pi{s}}\over{2}})}\over{s+1}}
\end{eqnarray}
with the summations subject to the same constraints (2.18).
The transformation law (2.16) is straightforward to generalize for any monomial
in the derivatives of $X$.
Namely, under $z\rightarrow{f(z)}$ we have

\begin{eqnarray}
:\partial^{n_1}X...\partial^{n_p}X:(z)\rightarrow
\nonumber \\
\sum_{q=1}^{\lbrack{p\over2}\rbrack}
\sum_{\lbrace{1...p}\rbrace\rightarrow{\lbrace}i_1...i_{2q};j_1...j_{p-2q}\rbrace}
\sum_{k_1=1}^{n_{j_1}}...\sum_{k_{p-2q}=1}^{n_{j_{p-2q}}}
S_{n_{i_{1}}|n_{i_{2}}}(f(z);z)...S_{n_{i_{2q-1}}|n_{i_{2q}}}(f;z)
\nonumber \\
B_{n_{j_1}|k_1}(f(z);z)...B_{n_{j_{p-2q}}|k_{p-2q}}(f(z);z):\partial^{k_1}X...
\partial^{k_{p-2q}}X:(f(z))
\end{eqnarray}
where
$\sum_{\lbrace{1...p}\rbrace\rightarrow{\lbrace}i_1...i_{2q};j_1...j_{p-2q}\rbrace}$
stands for the summation over the permutations
$\lbrace{1...p}\rbrace\rightarrow{\lbrace}i_1...i_{2q};j_1...j_{p-2q}\rbrace$
such that $i_1\neq{i_2}...\neq{i_{2q}}\neq{j_1}...\neq{j_{p-2q}}$;
$1\leq{i_k}\leq{p};1\leq{j_k}\leq{p}$ and $i_{2k-1}\leq{i_{2k}}$
(the last constraint is imposed in order to ensure that the redundant
combinations of  
Schwarzians $S_{n_i|n_j}$ do not appear in the permutations).
 
In what follows, we will be particularly interested in the terms
with $p=2q$ in the sum (2.21) that contain no operators but are just the numbers
only depending on $f(z)$. We shall call these terms
{\bf pure Schwarzian contributions}, and they will be play an 
important role in the calculations below. To simplify
the notations, it is convenient to write
\begin{eqnarray}
S_{n_1...n_p}(f(z);z)=
\sum_{\lbrace{1...p}\rbrace\rightarrow{\lbrace}i_1...i_{p}\rbrace}
S_{n_{i_{1}}|n_{i_{2}}}(f(z);z)...S_{n_{i_{p-1}}|n_{i_{p}}}(f(z);z)
\end{eqnarray}
with the summation over permutations of $1....p$ defined as above.
We are now prepared to return to the conformal transformation (2.15)
of $<<Q\Psi_0,\Psi>>$.
First, consider the transformation of $I\circ\Psi$ located at infinity.
Note that , in general $\Psi$ has the form similar to (2.6), except that,
generally speaking, $\alpha_{n_1...n_p}$-coefficients may depend on $X$.
According to the transformation formula (2.21), each term
in $\Psi$ gets multiplied by $e^{ihz}|_{z\rightarrow\infty}$ with
$h\geq{p-2q}$. Therefore all the contributions,
except for the one with $p=2q$ (that is, the pure Schwarzian contribution
$S_{n_1...n_p}$)
 are exponentially dumped and vanish identically
at infinity. So for any positive $N=n_1+...+n_p$ the only surviving 
part in any
component of $\Psi$ upon the conformal transformation (2.15) is the pure 
Schwarzian (which is constant, given by sum of combinations
of the products involving Stirling numbers according to (2.20)).
The only possible exception to it is the component with $N=0$
which, in principle, also may be present in $\Psi$.
This component is just a function of $X$ with no derivatives having the form $:f(X):$. But such a 
component a priori does not contribute to the contractions
with $\Psi_0$ in the correlator (note that $\Psi_0$ by construction contains no $N=0$ terms).
To see this, it is convenient to apply the conformal transformation $I(z)$ to the correlator
$<<Q\Psi_0(0)I{\circ}(:X^n:(0))>>$ for any $n$, taking :$X^n$: from infinity to zero and $\Psi_0$ at 0
to ${\tilde{Q\Psi_0}}$ at infinity, with ${\tilde{Q\Psi_0}}$ having the same form (2.6) as $Q\Psi_0$,
but with some new coefficients ${\tilde{\alpha}}_{n_1...n_p}$, straightforward to determine
from the conformal transformation. Note that $X^n$ doesn't change
as the resulting  conformal transformation applied to it, $I\circ{I}$ , is an identity. Then, using the translational invariance, take
$f(X)$ to $z=-\pi$, and apply another transformation $f(z)=e^{iz}$ to the correlator

$$<X^n(-{\pi\over2})(I\circ{{{{Q\Psi}}_0}})(\infty)>.$$ 

Similarly to what we explained before, 
only the pure Schwarzian terms remain out  of ${\tilde{\psi}}_0$ upon the transformation, implying
that the entire correlator is proportional to the pure Schwarzian factor of $X^n$ which does not contract.
But this factor is proportional to 

$$(S_{0|0}(f(z);z))^{n\over{2}}|_{f(z)=e^{iz};z={-{\pi\over2}}}$$
where $S_{0|0}={\log{(f^\prime(z))}}$, i.e. vanishes at $z=-{{\pi\over2}}$.
This shows that the only possible string field component of $I\circ\Psi$,
that does not vanish under $f(z)=e^{iz}$, except for the pure Schwarzian part,
does not contribute to the correlator $<<Q\Psi_0I\circ{\Psi}>>$.
 But then, since only the pure Schwarzian (non-contracting)
terms of $\Psi$ 
contribute to the correlator, the same is true for $Q\Psi_0$; 
therefore we conclude that the correlator $<Q\Psi_0(0)I\circ(\Psi(0))>$
evaluated on the singularoid has the form:

\begin{eqnarray}
<<Q\Psi_0\Psi>>=G_\Psi\sum_{N=1}^\infty\sum_{p=1}^{N}\sum_{N|n_1...n_p}\alpha_{n_1...n_p}S_{n_1...n_p}
\end{eqnarray}
where $G_\psi$ is some constant which only depends
on particulars of $\Psi$ and independent on $\Psi_0$.
The coefficients $\alpha_{n_1...n_p}$ are now to be chosen so that the correlator involving 
the summation over $N$ vanishes. At the first glance, this doesn't seem to be a simple problem because
of the complexity of $S_{n_1...n_p}$-factors involving cumbersome summations over
products of  generalized
Schwarzians.
There is, however, a simplification trick making it possible to deduce $S_{n_1...n_p}$
(as previously, we consider $p$ even).
Consider the correlator of $Q\Psi_0$ with ${1\over{p!}}:I\circ(\partial{X})^p:$ 
(multiplied by the c-ghost, as usual) in OSFT for some $p$.
The relevant terms in the part of  $<<Q\Psi_0,(\partial{X})^p>>$
for a given $N$  are
\begin{eqnarray}
{1\over{p!}}\sum_{N|n_1...n_p}<{{\partial^{n_1}X...\partial^{n_p}X(0)}\over{n_1!..n_p!}}I(\circ(\partial{X})^p(0))>
\nonumber \\
={\lim_{w\rightarrow{\infty}}}
{1\over{p!}}\sum_{N|n_1...n_p}<{{\partial^{n_1}X...\partial^{n_p}X(0)}\over{n_1!...n_p!}}w^{2p}(\partial{X})^p(w))>
(U_{0}(w))
\end{eqnarray}
where $U_0(w)$ is the overlap factor accounting for
for the correlator change  as a result of the integration of conformal Ward identities 
(note that the correlator (2.24), computed naively without this factor would have been proportional to 
to $\sim{w^{p-N}}$, i.e. would have vanished, as the $w^{2p}$-factor due to the conformal transformation
of $(\partial{X})^p$ by $I$ would have been multiplied by $w^{-N-p}$  as a result of the contractions). 

In our case, this factor is not difficult to compute explicitly.
Infinitezimally, it is given by the integral
\begin{eqnarray}
{1\over{p!}}\delta_\epsilon
\sum_{N|n_1...n_p}<{{\partial^{n_1}X...\partial^{n_p}X(0)}\over{n_1!..n_p!}}I\circ((\partial{X})^p(0))>
\nonumber \\
=-\lbrack{1\over{2p!}}\oint{{dz}\over{2i\pi}}\epsilon(z)\partial{X}\partial{X}(z);
\sum_{N|n_1...n_p}<{{\partial^{n_1}X...\partial^{n_p}X(\xi)}\over{n_1!...n_p!}}
(\partial{X})^p(w)>\rbrack|_{overlap;\xi=0,w\rightarrow\infty}
\nonumber \\
=
\sum_{N|n_1...n_p}\sum_{j=1}^p{1\over{(p-1)!n_1!...n_{j-1}!n_{j+1}!...n_p!}}
\nonumber \\
\times
<\partial^{n_1}X...\partial^{n_{j-1}}X\partial^{n_{j+1}}X...\partial^{n_p}X(\xi)
(\partial{X(w)})^{p-1}>|_{\xi=0,w\rightarrow\infty}
\nonumber \\
\times
\oint{{dz}\over{2i\pi}}{{\epsilon(z)}\over{(z-\xi)^{n_j+1}(z-w)^2}}
\end{eqnarray}
with one of the $\partial{X}$'s in the stress tensor $T(z)$ acting on the operator at $\xi$
and another on the operator at $w$ (i.e. the infinitezimal overlap transformation
gives the change of the entire correlator under the conformal transformation 
excluding the contributions due to infinitezimal conformal transformations of the vertex operators themselves).
The integral over $z$ is straightforward to evaluate, however, since the conformal transformation by
$I(z)$ only acts on the second operator  in $<<Q\Psi_0;\Psi>>=<Q\Psi_0(0){I}\circ\Psi(\infty)>$
 (in our case, $(\partial{X})^p$), only the pole at $w$ contributes to the overlap
function , so the $z$-integral's contribution to the infinitezimal overlap transformation is
\begin{eqnarray}
\sum_{j=1}^p\partial_{w}\lbrack{{\epsilon(w)}\over{(w-\xi)^{n_j+1}}}\rbrack=
\sum_{j=1}^p{{\partial\epsilon(w)}\over{(w-\xi)^{n+1}}}-(n_j+1){{\epsilon(w)}\over{(w-\xi)^{n_j+2}}}
\end{eqnarray}
This is easily integrated to give the finite transformation, i.e. the overlap function for $I(z)$:
\begin{eqnarray}
U_0(w)=\prod_{j=1}^p{{{{dI}\over{dz}}|_{z=w}}\over{(I(w)-I(\xi))^{n_j+1}}}|_{\xi=0;w\rightarrow\infty}
=w^{N-p}
\end{eqnarray}
Multiplying by the overlap function thus precisely cancels the vanishing $w^{p-N}$-factor discussed above,
keeping the correlator finite and relating the correlators before and after the conformal transformations.
The correlator is then easy to compute, as each given combination $n_1....n_p$, divided by $p!$, 
contributes exactly 1 to the correlator. Therefore the overall correlator simply equals the number 
of such combinations, i.e. the number of partitions $\lambda(N|p)$ of number $N$ with the length $p$:

\begin{eqnarray}
\sum_{N|n_1...n_p}<{{\partial^{n_1}X...\partial^{n_p}X(0)}\over{n_1!..n_p!}}(I\circ(\partial{X})^p)(\infty))>
=\lambda(N|p)
\end{eqnarray}
Next, apply the conformal transformation $f(z)=e^{iz}$ to the correlator (2.28).
Similarly to the explained above, the correlator computed on singularoid  is contributed by
the pure Schwarzian terms only with the overlap function computed to be
\begin{eqnarray}
U_0(w)={{p!w^{-2p}}\over{((p-1)!!)^2(S_{1|1}(e^{iz};z))^p}}+O(e^{iw})
\end{eqnarray}
where the Schwarzian of the exponential transformation $S_{1|1}(e^{iz};z)$ is simply ${1\over{12}}$.
Therefore the  correlator (2.28) computed on the singularoid, is given by
\begin{eqnarray}
{{\sum_{N|n_1...n_p}S_{n_1...n_p}}\over{(p-1)!!(S_{1|1}(e^{iz};z))^{p\over2}}}
\end{eqnarray}
and we deduce
\begin{eqnarray}
{\sum_{N|n_1...n_p}S_{n_1...n_p}}={\lambda(N|p)(p-1)!!(S_{1|1}(e^{iz};z))^{p\over2}}
\end{eqnarray}
 This identity particularly expresses the number of partitions of  the length $p$
in terms of summation (2.19), (2.20), (2.30) over Stirling numbers of the second kind.
Given (2.31), it is now straightforward to get the OSFT analytic solution of the form (2.6) for $\Psi_0$.
First of all, it is necessary to pick $\alpha_{n_1...n_p}=0$ for any $p$ odd, since
for the odd $p$ values the factorization (2.23) of the OSFT correlator $<<Q\Psi_0;\Psi>>$
doesn't appear to exist. For even $p$, writing $p=2k$, the solution for $\Psi_0$ is

\begin{eqnarray}
\Psi_0=c\sum_{N=2}^\infty{{\beta(N)}\over{\lambda(N)}}\sum_{k=1}^{\lbrack{N\over2}\rbrack}
\sum_{N|n_1...n_{2k}}\prod_{j=1}^{2k}{{\partial^{(n_j)}X}\over{{n_j!\sqrt{12}}}}
\nonumber \\
\lambda(N)\equiv\sum_{k=1}^{\lbrack{N\over2}\rbrack}(2k-1)!!\lambda(N|2k)
\nonumber \\
\beta(N)={{(N-1)\zeta(3)-\zeta(2)}\over{(N-1)^4}}
\nonumber \\
\end{eqnarray}
where $\zeta$ is the Riemann's zeta-function
and the $\lambda(N)$ coefficients are sums over the partitions of $N$ with even lengths
$2k$, weighted with $(2k-1)!!$.
Note that $\zeta(2)={{\pi^2}\over{6}}\approx{1.64}$ and $\zeta(3)\approx{1.2}$ is the Apery's constant.
Indeed, it is now easy to check that, with the string field given by (2.32)
one has
\begin{eqnarray}
<<Q\Psi_0,\Psi>>=G_\psi\sum_{N=2}^\infty({{\zeta(3)}\over{(N-1)^2}}-{{\zeta(2)}\over{(N-1)^3}})
=G_\psi(\zeta(3)\zeta(2)-\zeta(2)\zeta(3))=0
\end{eqnarray}
Note that all the $\beta(N)$ coefficients are positive, except for the $N=2$; in particular, we will discuss below the implications
of that for the entanglement of the bosonic string states.
This OSFT solution is straightforward to generalize to $D$ space-time dimensions;
one just has to take the product of $D$ copies  of $\Psi_0$:
\begin{eqnarray}
\Psi_0^{(D)}=c\prod_{m=1}^D\Psi_0^{(m)}
\nonumber \\
\Psi_0^{(m)}=\Psi_0(X\rightarrow{X_m})
\end{eqnarray}
($X$ is replaced with $X_m$ with the $c$-ghost factor removed)

The solution (2.32) has a structure quite different from the class of  the elementary solutions (2.4).
The summation over $N$ is essentially the summation over space-time spin values coinciding with
conformal dimensions of the string field's components; the components
 with different $k$ with $N$ fixed could then be understood as Stueckelberg terms for a given spin $N$.
Clearly, unlike the elementary solutions (2.4) defining wavefunctions of pure states,with given spins and masses,
 the solution (2.32) - (2.34) sums
over the ensemble of the states with different spins and masses , with the coefficients defining the reduced density
matrix of a certain subsystem.
As our solution carries he $b-c$ ghost number $1$( and in fact can be extended to superstring theory
with no coupling to the $\beta-\gamma$  ghost system), it belongs to the same ghost sector as the generators
of Poincare isometries in space-time.
It is therefore natural to identify  the solution (2.32) - (2.34) with the reduced density matrix of the subsystem
of the lower spin 1  entangled with tower of higher spins in open string theory, with the 
terms at a given $N$ corresponding to contribution from the spin $N$ subsystem to the entanglement.
In the next section we will give a more systematic explanation for such an identification;
it appears that the formalism of RNS superstring theory is more convenient for that; in particular
it makes it  far easier technically (in comparison with bosonic theory) to analyze the entanglement
of higher spin subsystems with systems incliuding all the spin ensembles.
Given the BRST cohomology solution described above, we can now compute the entanglement entropy
associated with the SFT solution (2.32)-(2.34). There is one subtlety though, that has to be pointed out.
We aim to express $\Psi_0$ as a sum over the ensemble of the pure states (each of them characterized by a certain mass and a spin),
with the summation coefficients defining the eigenvalues of the reduced density matrix. All of these coefficients must be positive
(since they correspond to classical probabilities). The coefficients that we computed are, on the other hand,
proportional to $\sim{{\beta(N)}\over{\lambda(N)}}$ and indeed are all positive, with the exception of 
first term   with $N=2$. 
Since  each $N$ represents the entanglement of spin 1 excitations with those of higher spin $N$,
to keep the density matrix Hermitian, it is sufficient to invert the sigh of the graviton's wavefunction 
(while keeping all the higher-spin wavefunctions invariant).
Writing
\begin{eqnarray}
{\tilde\Psi}_0=c\sum_{N=3}^\infty{{\beta(N)}\over{\lambda(N)}}\varphi_N
\end{eqnarray}
where by definition
\begin{eqnarray}
\varphi_N=\sum_{k=1}^{\lbrack{N\over2}\rbrack}
\sum_{N|n_1...n_{2k}}\prod_{j=1}^{2k}{{\partial^{(n_j)}X}\over{{n_j!\sqrt{12}}}}
\end{eqnarray}
we see that in $D$ space-time dimensions the SFT solution $\Psi_0^{(D)}$
can be written as
\begin{eqnarray}
\Psi_0^{(D)}=c\sum_{N_1,...,N_D\geq{3}}\rho_{N_1...N_D}\varphi_{N_1}^{(1)}...
\varphi_{N_D}^{(D)}
\end{eqnarray}
($\varphi_N^{(m)}$ is obtained from $\varphi_N$ by replacing
$X\rightarrow{X^m}$; $m=1,...,D$)

with the products of $\varphi_{N_j}$'s defining the ensemble of states
for the reduced density matrix of the lower-spin subsystem, with
$\rho_{N_1...N_D}$ defining the entanglement probabilities of this subsystem with the higher spins.
Note that the factors of $n_j!\sqrt{S_{1|1}}=n_j!{\sqrt{12}}$ appearing in $\varphi_{N}$ can be absorbed
by rescaling $\partial^{n_j}X$'s  in the products. Such a rescaling only affects an overall normalization
constant for the density matrix (call it $\lambda_0$), which in any case can be fixed from the 
condition:
\begin{eqnarray}
Tr\rho=\sum_{N_1,...,N_D}\rho_{N_1...N_D}=1
\end{eqnarray}
In particular, in $D=1$ the normalization condition reads
\begin{eqnarray}
\lambda_0^{-1}(\zeta(2)-\zeta(3)+\sum_{N=3}^\infty\beta(N)(\lambda(N))^{-1})=1
\end{eqnarray}
so SFT solution (2.32) must be divided by
\begin{eqnarray}
\lambda_0=\zeta(2)-\zeta(3)+\sum_{N=3}^\infty{\beta(N)}\lambda^{-1}(N)\equiv
\sum_{N=2}^\infty{|\beta(N)|}\lambda^{-1}(N)
\end{eqnarray}
to give  the normalized density matrix
(note that the summation over $N$ converges fast since the partition numbers
$\lambda(N)$ grow exponentially with $N$).
Accordingly, in $D$ dimensions the solution $\Psi_0^{(D)}$ (2.37) is to be divided
by $\lambda_0^D$ to ensure the  correct normalization.
That said, the entanglement entropy for the solution (2.37) is
\begin{eqnarray}
S_{ent}^{spin 1|all spins}
=D{\log{\lambda_0}}+{D\over{\lambda_0}}(\sum_{N=3}^\infty{{\beta(N)}\over{\lambda(N)}}
{\log{({{\lambda(N)}\over{\beta(N)}})}}-(\zeta(2)-\zeta(3))log(\zeta(2)-\zeta(3)))
\end{eqnarray}

The series in $N$ again converges fast as $\lambda(N)$ grows exponentially. It is tempting
to assume that each term in the summation
represents contribution of spin $N$ to the entanglement.
This concludes the computation of the entanglement entropy of the lower spins  as a subsystem
of the higher-spin system.

In the next section we shall generalize this computation to obtain the entanglement entropy
of a given spin $s$ subsystem, as a part of the entire higher-spin system. This will also provide an
additional explanation for the interpretation of the entropy (2.41) as the one for  the entanglement
of the lower-spin subsystem, discussed above.

\section{\bf Entanglement of spin $s$ subsystems: general case}

In this section we  will generalize the main result of the previous one and compute
the entanglement entropy of any spin $s$ subsystem.
For reasons that will become clear below,
it appears that the framework of RNS superstring theory is more convenient for this purpose,
compared to bosonic string theory. The action for the RNS superstring theory in superconformal gauge is
\begin{eqnarray}
S\sim{\int}{d^2z}\lbrack{-}{1\over2}\partial{X^m}\bar\partial{X_m}-{1\over2}\bar\partial\psi^m\psi_m-{1\over2}
\partial{\bar\psi}^m{\bar\psi}_m
\nonumber \\
+b\bar\partial{c}+{\bar{b}}\partial\bar{c}+\beta\bar\partial\gamma+{\bar\beta}\partial
{\bar\gamma}\rbrack+S_{Liouville},
\end{eqnarray}
the BRST charge (ignoring the Liouville terms) is now
\begin{eqnarray}
Q=\oint{{dz}\over{2i\pi}}\lbrack{cT-b{c\partial{c}}-{1\over2}\gamma\psi^{m}\partial{X_m}-
{1\over4}b\gamma^2}\rbrack
\end{eqnarray}
where $T$ is the full matter$+$ghost stress-energy tensor and, as before,
we are searching for the solutions of the linearized SFT equation $Q\Psi_0=0$, equivalent
to finding $\Psi_0$ such that $<<Q\Psi_0,\Psi>>=0$ for any $\Psi$ (subject to the gauge constraint $b_0\Psi=0$).
The bosonization relations for the ghost fields are, as usual

\begin{eqnarray}
c=e^\sigma,b=e^{-\sigma}
\nonumber \\
\gamma=e^{\phi-\chi},\beta=e^{\chi-\phi}\partial\chi
\end{eqnarray}

First of all, it is straightforward to extend the solution found in the previous section to
superstring theory. For simplicity, consider the number $D$ of space-time dimensions even.
Then, bosonize the RNS fermions according to

\begin{eqnarray}
\psi_{2j-1}\pm\psi_{2j}={\sqrt{2}}e^{\pm{i\varphi_j}}(j=1,...,{D\over2})
\end{eqnarray}
implying
\begin{eqnarray}
:\psi_{2j-1}\psi_{2j}:=-\partial\varphi_j
\end{eqnarray}

Since the stress tensor for $\psi_m$:
$$T_\psi=-{1\over2}\partial\psi^m\psi_m=-{1\over2}\partial\varphi_j\partial\varphi^j$$
doesn't have a background charge, conformal transformations of products of $\varphi$ derivatives
involve the generalized Schwarzians identical to those appearing in (2.18) for the $X$-fields.
Therefore the corresponding solution for $\Psi_0$ in superstring theory is simply
\begin{eqnarray}
\Psi_0^{(D)}=c\prod_{m=1}^{{3D}\over2}\Psi_0^{(m)}
\nonumber \\
\Psi_0^{(m)}=\Psi_0(X\rightarrow{X_m});m=1,...D
\nonumber \\
\Psi_0^{(m)}=\Psi_0(X\rightarrow{\varphi_{m-D}});m=D+1,...,{3D\over2}
\end{eqnarray}
($X$ is replaced with $X_m$ or $\varphi_{m-D}$ with the $c$-ghost factor
removed)
and the entanglement entropy is
\begin{eqnarray}
S_{spin 1|all spins}
={{3D}\over2}{\log{\lambda_0}}+{{3D}\over{2\lambda_0}}\sum_{N=2}^\infty{{|\beta(N)|}\over{\lambda(N)}}{\log{({{\lambda(N)}\over{|\beta(N)|}})}}
\end{eqnarray}
where $\Psi_0$ has the same form as in (2.32) with $X$ replaced with $X_m$ for $m=1,...D$ in $\Psi_0^{(m)}$
and with $\varphi_1,...\varphi_{{D\over2}}$ in the remaining ${D\over2}$ factors. The  entropy is then 
obtained from the one in the bosonic theory simply  by replacing $D\rightarrow{{3D}\over2}$.
Now let us consider the entanglement of a given spin $s\geq{3}$ system, regarded as the subsystem of 
string excitations with all the  spins.
As previously, the first step is to determine the appropriate solution in linearized string field theory.
To identify the structure of the solution we are looking for, it is useful to recall the general relation 
between the SFT solutions and the physical  vertex operators and currents in string theory.
For example, consider the Schnabl's solution \cite{schnablf, schnabls, schnablt} for nonperturbative tachyonic vacuum that was used to prove Sen's conjecture \cite{senf, sens, sent}.
This is the pure ghost solution, with the ghost number $+1$. Since at zero momentum the tachyon vertex operator
is just a $c$-ghost, the solution found by Schnabl was identified, based on its ghost-matter structure,
to the  nonperturbative tachyonic vacuum, defined by acting with this solution on the initial string vacuum state.
In the similar spirit,  we have identified the string field theory solutions (2.32) (carrying $b-c$ ghost number 1 and $\beta-\gamma$
ghost number zero, just as Poincare generators at unintegrated $b-c$ picture) with the reduced density matrix of the spin 1
system, considered as a subsystem of string modes with all the spins. The above arguments make it quite clear what type of the
solutions we should be looking for. Namely, to describe the reduced  density matrix of the subsystem with a given spin $s$,
we have to search for the SFT solutions in the superconformal  ghost sector containing the currents
 - the primaries of dimension one  integrated over the worldsheet's boundary (or multiplied by the $c$-ghost at unintergrated picture).
However, not all such operators generate authentic space-time symmetries, with some of them being BRST exact and some
 being the picture-changing transformations of operators with lower values of the ghost numbers.
In fact, the spin $s$ operators we need are the superconformal ghost number  $-s$ dimension zero primaries
satisfying the constraints \cite{selfframe, selfframes}

\bea
{\lbrace}Q,V^{(-s)}\rbrace=0
\nonumber \\
V^{(-s)}\neq{\lbrack}Q,...\rbrack
\nonumber \\
:\Gamma{V^{(-s)}}:=0
\eea

where

\bea
\Gamma=-{1\over2}e^\phi\psi_m\partial{X^m}-{1\over4}be^{2\phi-\chi}(\partial\chi+\partial\sigma)+ce^\chi\partial\chi
\eea
is the picture-changing operator for $\beta-\gamma$ pictures
or, in the dual positive $s-2$-picture:
\bea
{\lbrace}Q,V^{(s-2)}\rbrace=0
\nonumber \\
V^{(s-2)}\neq{\lbrack}Q,...\rbrack
\nonumber \\
:\Gamma^{-1}{V^{(s-2)}}:=0
\eea

Sets of  operators with such properties define the dual negative and positive
ghost hohomologies $H_{-s}\sim{H_{s-2}}$ \cite{selfframe}.

In the manifest form such operators can be constructed as follows:

Take a massless spin 3 operator, the element of $H_{-3}$, given by

\bea
V^{(-3)}=\Omega^{n_1n_2n_3}(p){\oint{{dz}}}e^{-3\phi}\psi_{n_1}\partial{X_{n_2}}\partial{X_{n_3}}e^{ipX}
\eea
 with the symmetric space-time spin 3  field $\Omega$ satisfying Fronsdal's on-shell constraints.
Consider the operator product of $2$ $V^{(-3)}$'s which is straightforward to calculate.
It is  straightforward to check that this product will have the form
\bea
V^{(-3)}V^{(-3)}\sim\partial{c}\sum_W{C_{w|v}^{(-3)}}{W^{(-3)}}+{C_{w|v}^{(-4)}}{W^{(-4)}}
\eea
where 
$C_{w|v}$ are the OPE structure constants with $W^{(-3)}$ and $W^{(-4)}$ being the vertex operators
from $H_{-3}$ and $H_{-4}$ respectively (with no operators from $H_{-5}$ and $H_{-6}$ , despite that
the right-hand side of the product has ghost number $-6$).
In particular, the ${W^{(-4)}}$ terms contain an operator 
which, after double picture-changing transformation from picture $-6$ to picture $-4$, takes the form
 $\sim{\oint{dz}}e^{-4\phi}\partial\psi_{(m_1}\psi_{m_2)}\partial{X^{n_1}}\partial{X^{n_2}}\partial{X^{n_3}}$
(times the structure constants multiplied by $\Omega^2$).
This operator is the vertex operator for the  two-row field $\Omega_{3|1}$ 
which, in Vasiliev's 
description, corresponds to the symmetric frame-like field of spin 4 and the structure constants
define the quadratic contribution of spin $3$ field to the $\beta$-function of the spin 4 field. 
This, in turn, produces a cubic $3-3-4$ vertex in the lower-energy effective action and
the appropriate term in the higher-spin algebra (that is, the spin $4$ term in the commutator
of two spin 3 currents).
For general $s$,  the operator algebra has the form:

\bea
:V^{(-s_1)}V^{(-s_2)}:\sim\partial{c}\sum_{s_3=|s_1-s_2|}^{s_1+s_2-2}\sum_W{C_{w|v}^{{s_1,s_2|s_3}}{W^{(-s_3)}}}
\eea
(the OPE coefficients $C_{w|v}^{{s_1,s_2|s_3}}$ vanish for $s_3=0,1,2$).
The general fusion rule for the cohomologies:
\bea 
H_{-s_1}{\otimes}H_{-s_2}\sim\sum_{s_3=|s_1-s_2|+\delta_{s_1s_2}}^{s_1+s_2-2}H_{-s_3}
\eea
 reproduces the structure of the higher-spin symmetry algebra with the structure constants 
generating the cubic couplings for the higher-spin frame-like fields in space-time.
All the above arguments altogether instruct us about the form of the SFT solution to search, in order to
describe the spin $s$ entanglement.
While still retaining the gauge condition $b_0\Psi=0$, it is appropriate to replace the 
constraint $\xi_0\Psi=0$ with the {\it cohomological gauge constraint}
\bea
:\Gamma\Psi:=0
\eea
for each negative ghost number sector and
\bea
:\Gamma^{-1}\Psi:=0
\eea
for each positive ghost number sector.
We will refer to the gauge choice (3.15), (3.16) as {\it cohomological gauge}.
This gauge choice is natural for our purposes since, with such a choice,
the SFT solutions at host number $-s$ can be clearly related  to
the reduced density matrices of the spin $s$ system;  with
other gauge choices, the ghost number $-s$ solutions would mix
contributions from different spins, with no obvious way to identify
 the entanglement.

First of all, the cohomological gauge imposes stringent limits on the 
possible number  of bosonized RNS fermions (  $\varphi$'s) in the solution.
That is, unlike the lower-spin $s<3$ SFT solution (2.32) 
with the number of of $\varphi$'s
unrestricted, the cohomological gauge restricts this number to $s-1$ at most, 
since
the OPE of $\partial\varphi$ with $\psi$ in $\Gamma$ has the structure
$\partial\varphi(z)\psi(w)\sim(z-w)^{-1}\psi(w)$ implying that any string
field containing product of more than $s-1$ derivatives of $\varphi$'s
would violate  cohomological condition. This precisely corresponds to the
number of the extra fields for a symmetric frame-like field of spin $s$ in Vasiliev's
formalism. This is again useful to compare with the structure of the higher-spin operators in the on-shell limit. Generally, the operators 
for the two-row $\Omega^{s|t}$ extra field with $t$ derivatives
($0\leq{t}\leq{s-1}$ contain  $t$ $\psi$-fields, with the $t=0$ 
field the only one being dynamical. In other words, the $\psi$-fields do 
not carry information about real physical degrees of freedom
in the cohomological gauge
 and should be excluded from the structure of the solution we are looking for.
Similarly, we shall ignore the components containing the derivatives of the $\phi$-ghost:
in the on-shell limit inclusion of the ghost derivatives in the vertex operator
effectively reduces the spin of the matter
part in space-time; for this reason the operators containing the ghost derivatives 
are related to the Stueckelberg-type terms (to ensure the overall BRST invariance of the operator)
and do not contribute to actual physical degrees of freedom.

That said, we shall search for the SFT solution in the cohomological gauge having the form:
\bea
\Psi_{s}=ce^{\chi-s\phi}\sum_{N_1=1}^\infty...\sum_{N_D=1}^\infty
\alpha(N_1,...,N_D)
\nonumber \\
\times
\sum_{k_1=1}^{\lbrack{{N_1}\over2}\rbrack}...\sum_{k_D=1}^{\lbrack{{N_D}\over2}\rbrack}
\sum^{(s-2)}_{N_1|n_1...n_{2k_1}}\prod_{j_1=1}^{2k_1}\partial^{n_{j_1}}X_1...\sum^{(s-2)}_{N_D|n_1...n_{2k_D}}
\partial^{n_{j_D}}X_D
\eea

where $\sum^{(m)}_{N|n_1...n_k}$ stands for the summation over the length k ordered partitions  
of $N=n_1+...+n_k$ such that $n_1\geq{n_2}...\geq{n_k}>0$ with values of the partition elements not bigger than $m$
(so $n_1\leq{m}$) 
Next, the BRST charge acting on  $\Psi_s$ gives:
\bea
Q\Psi_s=\partial{c}ce^{\chi-s\phi}\sum_{N_1=1}^\infty\sum_{R_1=1}^\infty...\sum_{N_D=1}^\infty\sum_{R_D=1}^\infty
\alpha(N_1,...N_D)(N_1+...N_D-{1\over2}s^2+s-1)
\nonumber \\
\times
\prod_{j=1}^D{{\Psi}}_s(X_j;N_j)+...
\eea
where we skipped the irrelevant terms, as was explained in the previous section.
From this we deduce
\bea
\alpha(N_1,...,N_D)=(N_1+...N_D-{1\over2}s^2+s-1)^{-1}
\prod_{j=1}^D{{\beta(N_j)}\over{\lambda^{(s-2)}(N_j)}}
\eea
where
\bea
{{\lambda^{(m)}}(N_j)\equiv\sum_{k=\lbrack{N_j\over{m}}\rbrack}^{\lbrack{N_j\over2}\rbrack}(2k-1)!!{{\lambda^{(m)}}}(N_j|2k)}
\end{eqnarray}
where
${{\lambda^{(m)}}}(N_j|2k)$ is the number of the length $2k$ partitions of $N_j=n_1+...+n_{2k};n_1\geq{n_2}\geq...\geq{n_{2k}>0}$
 with all the elements of the partition being not greater than $m$ or, in other words,
\bea
n_1\leq{m}
\eea
The remaining steps are identical to those described in the previous section.
With the conformal transformation
$z\rightarrow{f(z)=e^{iz}}$
we reduce the SFT correlator
$$<<Q\Psi_s;\eta_0\Psi>>=<Q\Psi_s(0)(I\circ(\eta_0\Psi))(\infty)>$$ to
pure Schwarzian contributions from $\eta_0\Psi$ at infinity and, consequently, $\Psi_s$.
The structure of the result is then identical to (2.32), given by series in combinations of generalized
Schwarzians for $\Psi_s$ (which orders are now restricted by the cohomological gauge constraints)
, multiplied by the factor depending on $\eta_0\Psi$ only (times the constant
given by the ghost part of the correlator) 
Then we compare it to the test correlator
$<<Q\Psi_s;\eta_0\Phi>>$ with the matter part of $\Phi$ given by
$\Phi_{matter}=\sum_{k_1,...k_D=1}^\infty\prod_{j=1}^D{1\over{k_j!}}:(\partial{X_j})^{k_j}:$. Using
the identity $<Q\Psi_s(0)I\circ(\Phi_{matter})(\infty)>=<Q\Psi_s(0)\eta_0\Phi_{matter}(1)>$
we express $<<Q\Psi_s;\eta_0\Phi>>$ in terms of the weighted numbers of restricted partitions $\lambda^{(s)}$ (3.20).
Finally, applying the transformation $f(z)=e^{iz}$ to the test correlator, we relate 
the sum over the combinations of the generalized Schwarzians to the restricted partition numbers.
It is then straightforward to check that, with the choice (3.19) of $\alpha$ the string field $\Psi_s$ satisfies
\bea
<<Q\Psi_s;\eta_0\Psi>>={\tilde{G}}_\Psi{\times}(\zeta(2)\zeta(3)-\zeta(3)\zeta(2))^D=0
\eea
for any $\Psi$ where, as before, 
${\tilde{G}}_\Psi$ is the factor that only depends on the string field $\Psi$ (but not on $\Psi_s$).
Thus the string field $\Psi_s$  (3.17), (3.19) defines the ghost number $-s$ linearized OSFT solution in the cohomological gauge,
which defines the reduced density matrix for a space-time spin $s$ subsystem in superstring theory.
With this density matrix, it is straightforward to obtain the entanglement entropy for the spin $s$ subsystem, 
with the result given by:
\bea
S_{ent}(s)=
{\log{{\tilde{\lambda}}^{(s)}}}-{1\over{\lambda^{(s)}}}\sum_{{\tilde{N}}=D;{\tilde{N}}\neq{{{s^2}\over2}-s}}^\infty
\rho_s({\tilde{N}}){\log{\rho_s({\tilde{N}})}}
\eea
where 
\bea
\rho_s({\tilde{N}})={1\over{{|\tilde{N}}+s-{{s^2}\over2}|}}\sum_{\lbrace{N_1},...,{N_D}:N_1+...+N_D={\tilde{N}}\rbrace}
{{|\beta(N_1)...\beta(N_D)|}\over{\lambda^{(s)}(N_1)...\lambda^{(s)}(N_D)}}
\eea
with the sum is taken over all positive values of $N_1,...,N_D$ satisfying
\be
N_1+...+N_D={\tilde{N}}
\ee
and
\bea
{\tilde{\lambda}}^{(s)}=\sum_{{\tilde{N}}=D;{\tilde{N}}\neq{{{s^2}\over2}-s}}^\infty\rho_s({\tilde{N}})
\eea
This  concludes the calculation of the entanglement for the spin $s$ subsystems.
It should be noted  that the SFT solutions (3.17), (3.19) are $GSO$-odd
and $GSO$-even for the odd and even spin values respectively. To preserve the algebraic structures of SFT, such as cyclicity
of the correlators, one can assign the internal Chan-Paton factors to the operators, e.g. by multiplying
 the GSO-even operators by $2\times{2}$ identity  matrix, GSO-odd operators by $\sigma_1$ Pauli matrix,
while multiplying $Q$ and $\eta_0$ by $\sigma_3$. Then, upon computing the SFT correlators, one has to take the trace
over the resulting $2\times{2}$ matrix.

In the following concluding section, we shall discuss some properties of our SFT solutions and the results
for the entanglement.

\section{\bf Conclusions and Discussion}

In this work we have calculated the entanglement entropies for the subsystems of spin $s$
excitations in string theory, using the solutions in linearized open string and superstring field theories.
Unlike the elementary solutions of the linearized OSFT that typically define the pure states (on-shell vertex operators
acting on the vacuum), the solutions that we find and analyze in this work
define reduced dencity matrices for various spin $s$ excitations and the related entanglements with other spins.
Despite the overall complexity  of the operators involved, the conformal transformations described in this paper
allow to express them in terms of series over generalized Schwarzians and, subsequently, to relate these series
to weighted partition numbers. The final answer for the entropies is remarkably simple - they all are 
expressed in terms of convergent series in the inverse partition numbers, with no the restrictions on the partition elements
for the lower-spin (spin 1) entanglement and with the values of the partitions  restricted by the spin value $s$
for the higher-spin entanglement.

The restrictions  on values of the partition elements for the higher-spin reduced density matrices and entanglements
 is the direct
consequence of the cohomological gauge condition,
necessary to  single out authentic higher-spin currents amidst higher ghost number string fields. 
This restriction  clearly reduces the number of relevant partitions $\lambda(N_j|2k)$ and hence
the number of weighted partitions ${{\lambda^{(s)}}}(N_j)$  entering the solution.
Since the density matrix elements are divided by the normalization factors ${\tilde{\lambda}}^{(s)}$
involving summations over inverse $\lambda^{(s)}(N_j)$, converging faster with $s$.
 this clearly implies  that the entanglement
entropy generally grows with $s$ for $s\geq{3}$.
Next,  our results for the entanglement entropies imply that the entanglements for any spin contain  universal contributions
which are purely logarithmic and have the form $\sim{{\log{{{\tilde{\lambda}}^{(s)}}}}}$
(where ${\tilde{\lambda}}^{(s)}$ is given by the series in terms of inverse weighted partition numbers with the
partition elements restricted by $s-2$ for $s{\geq}3$ and with no restrictions for the lower spins.
These purely logarithmic contributions  are collective in a sense that they can't be viewed as 
sums of individual contributions from different spins to the entanglement (unlike the terms linear
in inverse ${\tilde{\lambda}}^{(s)}$ in (3.23). 
The structure  of these contributions hints at their possible interpretation: these terms  represent the 
entanglement swappings between spin $s$ subsystems and 
the string vacuum, representing nonlocality of time in string theory, reminiscent of the entanglement
between non-coexisting photons that has been observed experimentally \cite{meg}.In string theory
context,
this swapping
is the entanglement between the spin $s$ excitations of a string and the vacuum state in the past.

In this paper we have calculated the lower (spin 1) entanglement in both bosonic OSFT and in
superstring field theory, while the calculation of the
 higher-spin entanglement  ($s{\geq}3$) was limited to superstring field theory only.
Calculating the higher-spin entanglement in bosonic  string field theory 
seems to be much harder to do because the analogue of the cohomological gauge, used
to identify the higher-spin density matrices in the set of higher ghost number SFT  string fields,
is far more complicated in the bosonic theory. This is because in 
bosonic theory the cohomological gauge has to be defined 
with respect to the $b-c$ picture changing operator
$Z=:b\delta(T):$ which is a highly nonlocal object (unlike $\Gamma$) due the delta-function of
the stress tensor (an object with conformal dimension $-2$). One needs to have a better understanding of the OPE structure of  
the $Z$-operators in order to extend our results to the bosonic theory.

In this work we limited ourselves to calculating the entanglement on the solutions of the linearized theory. 
It would be obviously extremely interesting and important to extend our results to the full   
interacting SFT, by the identifying the reduced density matrix type solutions. 
Although finding analytic solutions in interacting SFT isn't simple in general,
we hope that the singularization method that we used in this work, can be extended
to the  interacting theory with some modifications, in order to obtain new classes of solutions. 
Given the background independence of string field theory, it can be  holographically related 
to very different quantum field theories and systems, such as holographic fluids and condensed
matter systems. With the interplays between quantum entanglement and concepts of string field theory,
mentioned in the beginning our work, our hope is that SFT will prove to be a new powerful framework
for computing the entanglement entropies in various systems and for our understanding of
quantum entanglement in general (including its relevance to the origin of space and time). 
We hope to address these questions in our future works.

\section{\bf Acknowledgements}

The author gratefully acknowledges the support of National Science Foundation of China (NSFC) under the project
11575119.

\end{document}